\documentstyle[aps,preprint,prl,psfig,amssymb]{revtex}

\draft
\tighten
\begin{document}
\bibliographystyle{prsty}

\title{The relaxation dynamics of a simple glass former confined
in a pore}
\author{Peter Scheidler,
Walter Kob, and Kurt Binder}
\address{Institute of Physics, Johannes Gutenberg-University,
Staudinger Weg 7, D--55099 Mainz, Germany}
\date{15. March, 2000}
\maketitle

\begin{abstract}
We use molecular dynamics computer simulations to investigate the
relaxation dynamics of a binary Lennard-Jones liquid confined in a narrow pore. We
find that the average dynamics is strongly influenced by the confinement
in that time correlation functions are much more stretched than in
the bulk. By investigating the dynamics of the particles as a function
of their distance from the wall, we can show that this stretching is due
to a strong dependence of the relaxation time on this distance, i.e. that
the dynamics is spatially very heterogeneous. In particular we find that
the typical relaxation time of the particles close to the wall is orders
of magnitude larger than the one of particles in the center of the pore.

\end{abstract}

\vspace{-0.3cm}
\narrowtext

Despite the remarkable progress that has been made in recent years in our
understanding of the dynamics of supercooled 
liquids~\cite{vigo_1998,goetze_1999}
there are still situations in which the nature of this dynamics is
very unclear. One example is the behavior of supercooled liquids
in restrictive geometries, such as in thin films or in narrow
pores~\cite{klafter_proc_1989,drake_proc_1997,grenoble_confit_2000}. 
The motivation to study
the dynamics of fluids in such geometries lies in their importance
as catalysts, flow through porous materials, or capillaries in
biological systems, to name a few.  Apart from their relevance in
various applications, this dynamics is also of considerable interest
for basic science since it might help us to gain a better understanding
of the dynamics of supercooled liquids and the glass transition in the
{\it bulk}, a subject which is, despite the progress mentioned above,
still a field of very active research.  In particular this type of study
might be used to investigate the existence of a diverging length scale
in supercooled systems.  E.g.~investigations of systems that show a
conventional ordering phenomena with a growing characteristic length scale,
i.e.~phase transitions, have
shown that, by perturbing the system by means of a wall and by monitoring
over which distance this perturbation vanishes, the size of the length scale
can be determined. Thus it should be possible to use this approach
also in the case of supercooled liquids~\cite{parisi}. 

Since at low temperatures the modes leading to the relaxation of
the system are generally assumed to be very cooperative, one might
argue that the confinement will supress such modes and therefore
lead to a dynamics which is slower
than the one of the bulk. On the other hand one could also think
that the interaction between the liquid and the wall of the host
material leads to an acceleration of the dynamics, at least close to
the walls. In order to decide what is actually
happening many experimentalists started to study such systems.
Using new nano- and meso scale materials with adjustable pore size,
such as Vycor glass and zeolites, and various techniques, such as
differential scanning calorimetry, dielectric relaxation measurements,
and solvation dynamics, the relaxation dynamics was investigated
as a function of pore size, type of liquid and host material, temperature,
etc.~\cite{warnock_1986,schueller_1994,pissis_1998,arndt_1997,mckenna_1996,richert_1996,forrest_1997,kremer_1999,wendt_1999}.
Surprisingly the obtained results do not give a clear answer at
all because for certain materials the glass transition temperature
$T_g$ in the confined system is found to be higher than the one of
the bulk system~\cite{schueller_1994}, whereas in other systems it
seems to be lower~\cite{pissis_1998,arndt_1997,mckenna_1996}. Sometimes it
is also observed that the confinement does not affect the value
of $T_g$ at all~\cite{richert_1996}. However, one has to bear in
mind that one of the main problems of these experiments is the proper
interpretation of the observed results. Although the topology of the
pores can presently be controlled reasonably well~\cite{goeltner_1997}, the
structure  of a single pore is never determined exactly. Furthermore it is
experimentally not really possible to distinguish between the influence of
the fluid-wall interaction, which can be the dominant 
effect~\cite{forrest_1997}, on the one hand from pure confinement effects
on the other hand. Therefore it has so far not been possible to obtain
a clear picture of the relaxation dynamics in confined systems.

Also computer simulations have been done in order to
gain some insight into the dynamics of liquids in restrictive
geometries~\cite{sappelt_1993,fehr_1995,baschnagel_1996,boeddeker_1999,nemeth_1999,gallo_2000a,gallo_2000b,varnik_2000,scheidler_2000}.
Since in simulations it is easy to control the details of the geometry
(size, topology) and also the interaction between confined fluid
and host material, this method is well suited to investigate this
dynamics. In addition it is possible to measure {\it directly} all
quantities of interest, since at any time the positions and velocities
of the particles are known, an advantage of this method which has been
made use of extensively in simulations of the bulk~\cite{kob_1999}. In
the past most simulations with restrictive geometries have been done
for planar geometries~\cite{fehr_1995,baschnagel_1996,boeddeker_1999,varnik_2000}, whereas
curved geometries, i.e. tubes or cavities, have been considered only
seldom~\cite{nemeth_1999,gallo_2000a,gallo_2000b}. 
In the present paper we therefore report
the results of a large scale molecular dynamics computer simulation in
which we investigated the dynamics of a simple glass former in a narrow
pore. In particular we will focus on the dependence of the relaxation
dynamics on temperature and the distance from the confining wall.

The system considered is a binary mixture of 80\% A and 20\% B particles
that have the same mass $m$ and which interact via a Lennard-Jones
potential of the form $V_{\alpha\beta}(r)=4\epsilon_{\alpha\beta}
[(\sigma_{\alpha\beta}/r)^{12}-(\sigma_{\alpha\beta} /r)^{6}]$,
with $\alpha,\beta \in\{{\rm A,B}\}$ and cut-off radii
$r_{\alpha\beta}^{\rm C}$=$2.5\cdot \sigma_{\alpha\beta}$. The parameters
$\epsilon_{\alpha\beta}$ and $\sigma_{\alpha\beta}$ are $\epsilon_{\rm
AA}=1.0$, $\sigma_{\rm AA}=1.0$, $\epsilon_{\rm AB}=1.5$, $\sigma_{\rm
AB}=0.8$, $\epsilon_{\rm BB}=0.5$, and $\sigma_{\rm BB}=0.88$. In the
following all results will be given in reduced units, i.e. length in
units of $\sigma_{\rm AA}$, energy in units of $\epsilon_{\rm AA}$ and
time in units of $(m\sigma_{\rm AA}^2/48\epsilon_{\rm AA})^{1/2}$. For
Argon these units correspond to a length of 3.4\AA, an energy of 120K$k_B$
and a time of $3\cdot10^{-13}$s. The fluid is confined in a cylindrical
tube with radius $\rho_{\rm T}=5.0$ and below we will compare its properties
with the ones in the bulk, using results that have been obtained
earlier~\cite{kob_1995}. In agreement with other computer simulations of
confined systems~\cite{fehr_1995,nemeth_1999,varnik_2000}, we find that a {\it smooth} wall
leads to a (in our case strong) layering of the confined liquid, i.e. its
structure is very different from the one in the bulk.
Since our goal is to investigate the effect of the confinement onto the
dynamical properties and not the effect due to a change of structure,
a significant change of the structural properties is not acceptable. In
order to avoid this latter possibility we chose the wall of the pore
to have a liquid-like structure similar to the one of the confined liquid.
For this purpose, we equilibrated a large bulk system at a temperature at
which it is slightly supercooled, $T=0.8$, and extracted a cylinder with
radius $\rho_{\rm T}+r_{\rm AA}^{\rm C}$.  Those particles that had a
distance $\rho$ from the center of the axis of the cylinder larger than
$\rho_{\rm T}$ were subsequently frozen at their position and constituted
the wall particles. The particles for which $\rho < \rho_{\rm T}$
constituted the fluid and were allowed to continue to move according
to Newton's laws using the same Lennard-Jones potential as interaction.
With this choice of the wall the static properties of the confined system
(density profile, radial distribution function) remain essentially
unchanged~\cite{scheidler_2000}, and therefore the changes in the dynamic
properties are only due to the confinement.

The time evolution of the system was calculated by solving the equations
of motion with the velocity form of the Verlet algorithm with a time step
of 0.01 at high ($T\ge 1.0$) and 0.02 at low ($T\le 0.8$) temperatures.
To improve the statistics of the results we simulated between 8 and 16
independent systems, each containing 1905 fluid particles and about 2300
wall particles.  The length of the tube was 20.137 which resulted in an
average particle density of 1.2, the same value as used in the earlier
simulations  of the bulk~\cite{kob_1995}.  The temperatures investigated
were $T=5.0$, 2.0, 1.0, 0.8, 0.7, 0.6, and 0.55.  The equilibration was done
by coupling the liquid periodically to a stochastic heat bath for a
time which was sufficiently long to equilibrate the system. All data
presented here were produced during a micro-canonical run at constant
energy and volume. During this run we observed that the particles of the
fluid are able to enter the wall to some extent. Therefore we defined
a ``penetration radius'' $\rho_{\rm p}$ as the distance beyond which it is
very unlikely to find a particle of the fluid, i.e. the probability is
less than $10^{-4}$. We found that at low temperatures, $T \leq 0.7$,
the value of $\rho_{\rm p}$ is essentially independent of temperature and
that for the A and B particles it is $5.5\pm 0.2$ and $6.1 \pm 0.2$,
respectively. Below we will see that this penetration radius, a {\it static}
quantity, is also relevant for the {\it dynamics} of the system.

\begin{figure}[h]
\psfig{figure=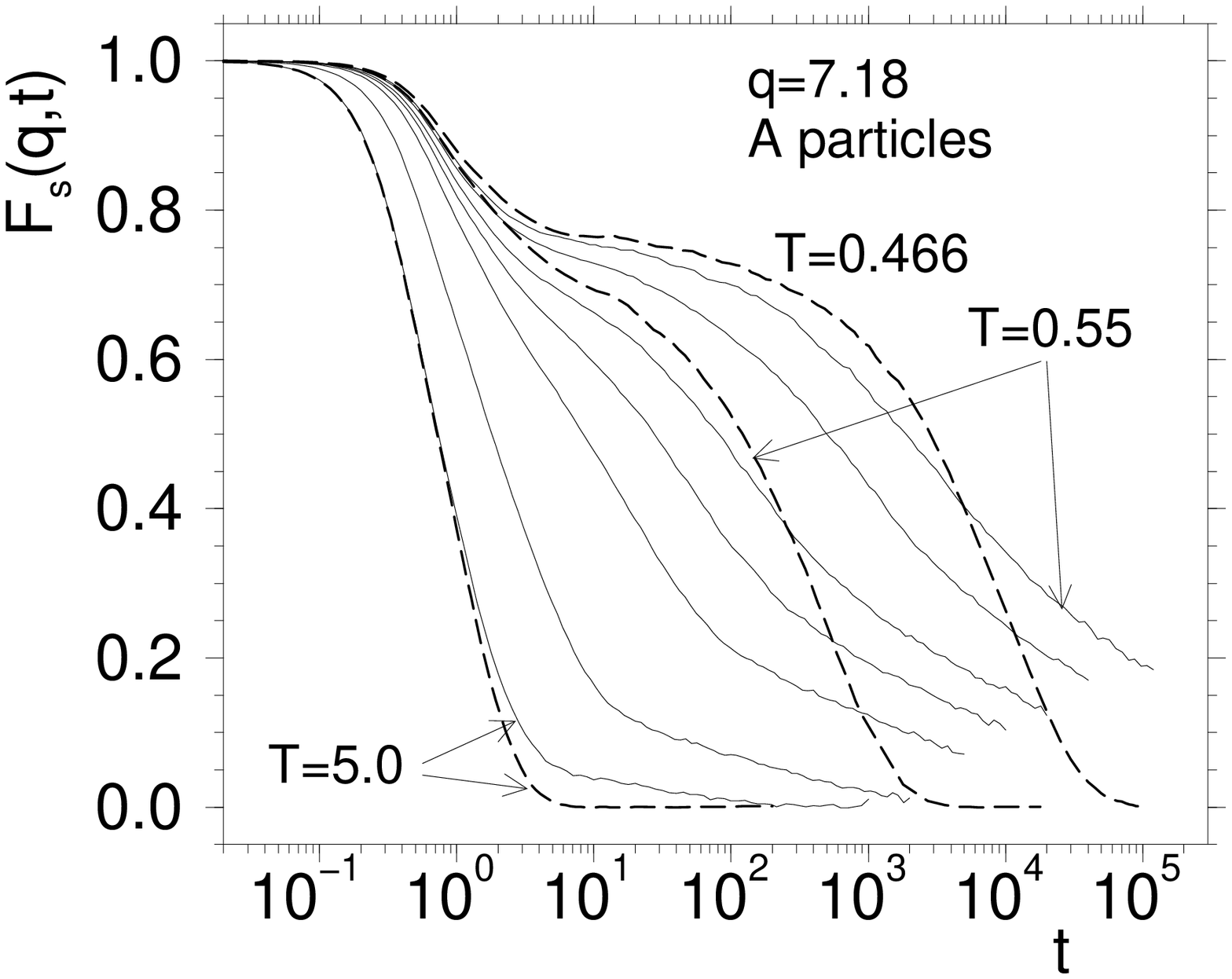,width=10cm,height=7.5cm}
\caption{Time dependence of the self intermediate scattering function
for the liquid in the tube (solid curves) for all temperatures investigated and
$q=7.18$ and in the bulk (dashed curves) at selected temperatures and 
$q=7.25$. 
}
\label{fig1}
\end{figure}

A quantity which in the past has been found to be useful to characterize
the dynamics of liquids in the bulk is the so-called self intermediate
scattering function $F_{\rm s}({\bf q},t)$ for wave-vector ${\bf q}$, which
is related to a density fluctuation $\rho_{\rm s}({\bf q},t)=\exp(i{\bf q}
\cdot {\bf r}_j(t))$ by $F_{\rm s}({\bf q},t)=\langle \rho_{\rm s}({\bf q},t)
\rho_{\rm s}(-{\bf q},0) \rangle$. (Here ${\bf r}_j(t)$ is the position
of particle $j$ and $\langle . \rangle$ is the canonical average.)
In Fig.~\ref{fig1} we show the time dependence of $F_{\rm s}({\bf q},t)$ for
all temperatures investigated. The wave-vector is along the axis of the
tube and has magnitude $q=7.18$, which is the location of the main peak
in the static structure factor $S(q)$~\cite{scheidler_2000,kob_1995}. As it
was the case in the bulk~\cite{kob_1995}, the correlation functions show
that with decreasing temperature the dynamics of the system slows down
rapidly and relaxes at low temperatures in two steps, an effect that is
related to the cage-effect. The details of this slowing down, as well
as the shape of the correlation function is, however, very different
from the bulk behavior. To see this we have included in the figure also
three bulk curves for $T=5.0$, 0.55, and 0.466 (bold dashed lines). We
recognize that, whereas the typical relaxation time of the function is
basically the same at $T=5.0$, a decrease of the temperature leads to
a slowing down of the dynamics which is much more pronounced in the restricted
system than in the bulk system (see the curves for $T=0.55$). In addition
to this also the shape of the curves depends on the geometry in that
even at the high temperature $T=5.0$ the two curves differ since the
one for the tube shows a small tail at long times ($t\geq 4$) that is
not present in the bulk curve. If the temperature is lowered this
tail becomes more pronounced and starts at higher and higher values of
$F_{\rm s}(q,t)$, whereas no such tail is observed in the bulk curves even at
the lowest temperatures. Thus the curve for the tube shows a much stronger
stretching than the one for the bulk, in agreement with the experimental 
findings of Refs.~\cite{schueller_1994,pissis_1998,arndt_1997,kremer_1999}.

In order to find the reason for this different dynamical behavior we
consider a generalization of the self intermediate scattering function
by defining the function $F_{\rm s}({\bf q},\rho,t)$ as

\begin{equation}
F_{\rm s}({\bf q},\rho,t)=\langle \exp[i{\bf q}\cdot ({\bf r}_j(t)-{\bf
r}_j(0))] \delta ({\bf r}_j(0)-\rho)\rangle \quad .
\label{eq1}
\end{equation}

\begin{figure}[h]
\psfig{figure=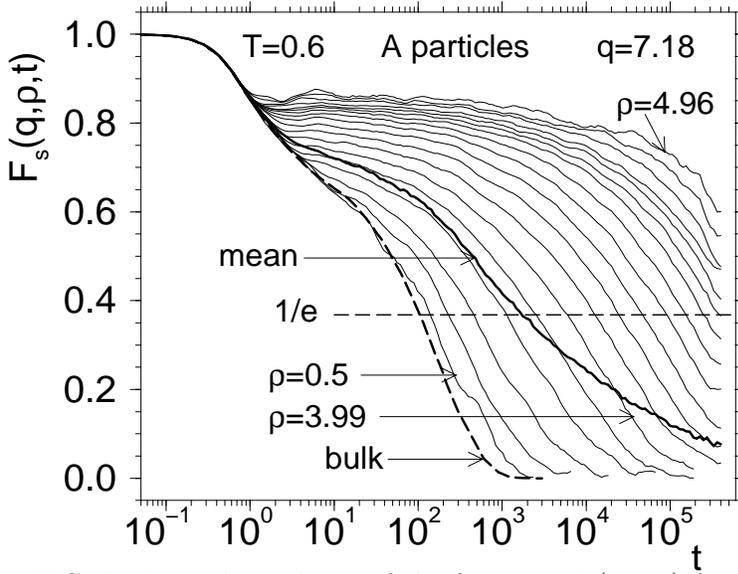,width=10cm,height=7.5cm}
\caption{Time dependence of the function $F_{\rm s}(q,\rho,t)$ for various distances
$\rho$ (thin solid curves). The bold solid curve and the dashed curves are 
the intermediate scattering function for the whole system and the bulk system,
respectively.}
\label{fig2}
\end{figure}

Thus this correlation function considers only particles that at time zero
had a distance $\rho$ from the central axis of the tube. By investigating
the $\rho$ dependence of this correlator it becomes hence possible
to understand which particles are moving relatively fast and which
ones relatively slow. In Fig.~\ref{fig2} we show the time dependence
of $F_{\rm s}({\bf q},\rho,t)$ at a low temperature. The direction of the
wave-vector is the same as in Fig.~\ref{fig1}, i.e. parallel to the
axis of the tube. The value $\rho_i$ for the distance $\rho$ is chosen
such that the values $(\rho_{\rm p}-\rho_i)^{-1}$ are spaced equidistantly,
where $\rho_{\rm p}$ is the penetration radius introduced above. (The reason
for this choice will become clear below.) Thus the leftmost curve
corresponds to $\rho=0.5$ and the rightmost one to $\rho=4.96$. From the
figure we recognize that the dynamics of the particles close to the
center of the tube is similar to the one of the bulk (shown in the
figure as bold dashed line).  However, for particles that are close to
the wall the dynamics is slowed down by several orders of magnitude. We
also note that although the time scale of the relaxation function depends
strongly on $\rho$, its {\it shape}, which could, e.g., be characterized by
the value of the Kohlrausch parameter $\beta$, seems to be independent
of it within the accuracy of our data and that this shape is
similar to the one found in the bulk. Since the relaxation curve of the
whole system is the weighted sum over the curves for the different $\rho$,
the resulting correlator is much more stretched than the one of the bulk,
in agreement with the observation from Fig.~\ref{fig1}.

\begin{figure}[h]
\psfig{figure=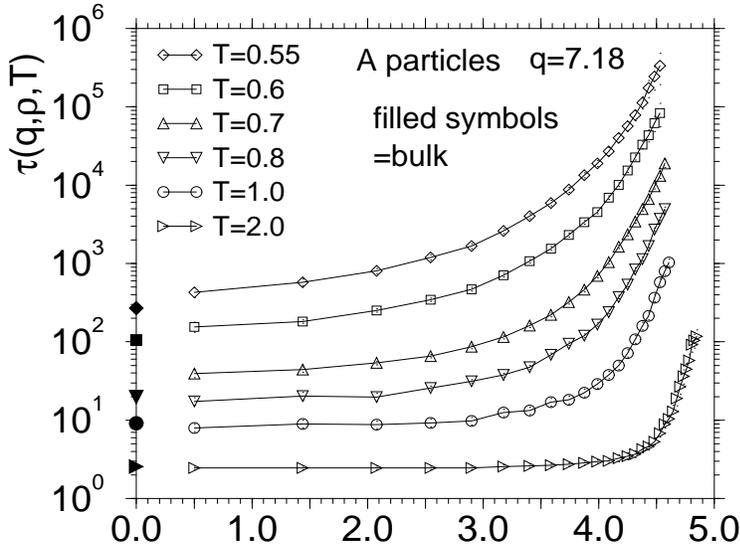,width=10cm,height=7.5cm}
\caption{$\rho$-dependence of the relaxation time for $T\leq 2.0$  
(curves with open symbols with size comparable to the error bars). 
The single symbols are the relaxation times for the
bulk system.}
\label{fig3}
\end{figure}

In order to investigate this $\rho$-dependence of the correlators in more
detail we
define the $\alpha$-relaxation time $\tau(q,\rho,T)$ by the time it
takes the correlation function to decay to $e^{-1}$ of its initial
value (see the horizontal dashed line in Fig.~\ref{fig2}). Such a
definition is reasonable since the shape of the correlation function
is essentially independent of $\rho$.  In Fig.~\ref{fig3} we plot this
$\rho$-dependence for the different temperatures (curves with open
symbols)~\cite{footnote}. For the sake of comparison we have included
in the figure also the relaxation times of the system in the bulk
(filled symbols). From this graph we see that at high temperatures the
relaxation time is independent of $\rho$, if $\rho$ is not too large,
and is the same as in the bulk. Only upon approach to the wall, $\rho
\geq 4.0$, $\tau(q,\rho,T)$ starts to increase significantly. With
decreasing temperature this crossover distance becomes smaller until
at the lowest temperatures the whole system is affected by the slowing
down due to the wall. Thus at these temperatures the dynamics
of all the particles is slower than the one of the bulk, which can be
recognized from the fact that an extrapolation of the curves towards
$\rho=0$ intercepts the vertical line at $\rho_{\rm p}^{-1}$ above 
the value for the bulk.

\begin{figure}[h]
\psfig{figure=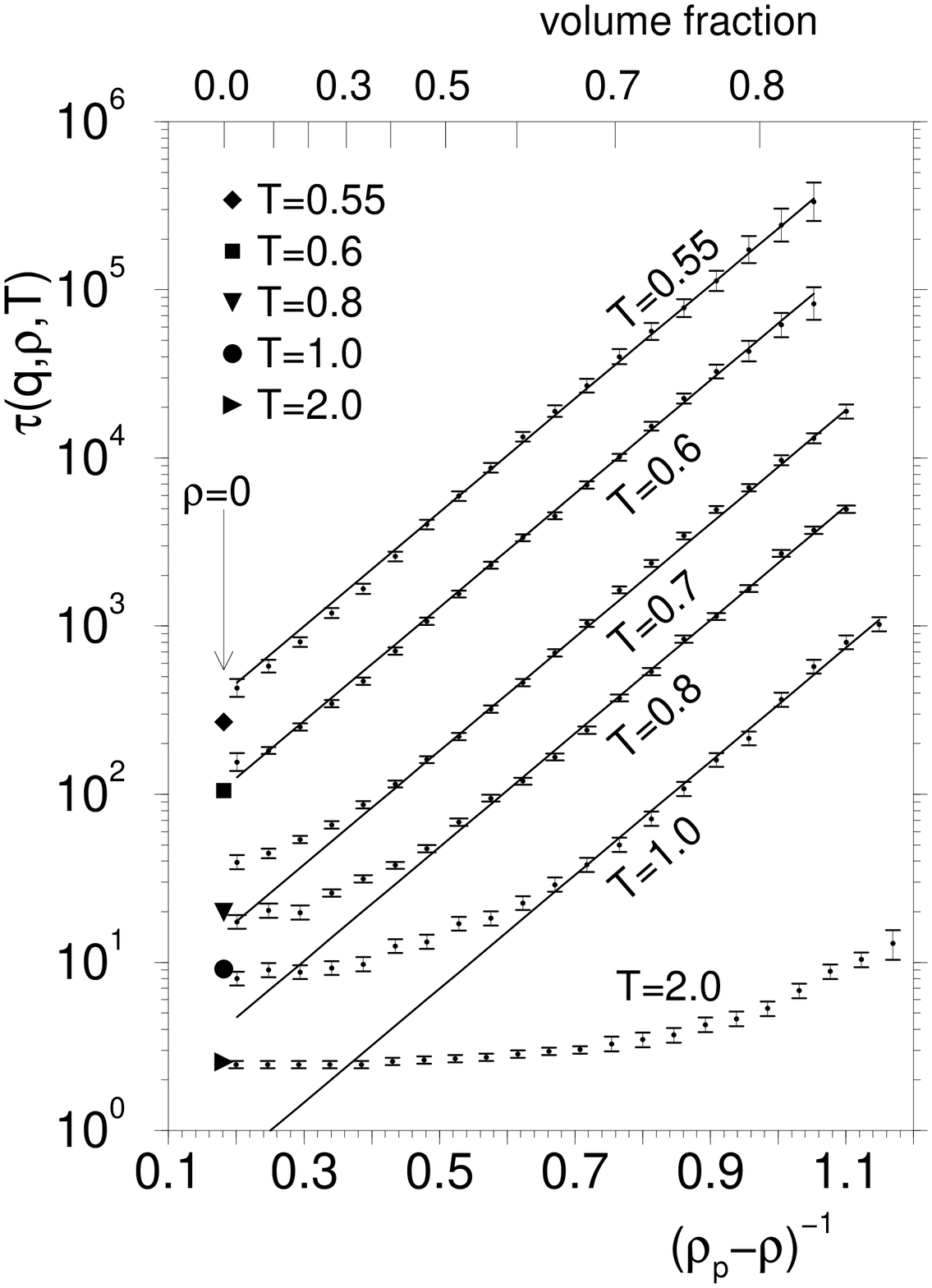,width=7.5cm,height=10cm}
\caption{Relaxation time as a function of $(\rho_{\rm p}-\rho)^{-1}$, where $\rho_{\rm
p}$ is
the penetration radius of the pore. The filled symbols are the relaxation times in
the bulk. The straight lines are fits with the functional form of
Eq.~(\protect\ref{eq2}) with $\Delta_q$ independent of $T$.}
\label{fig4}
\end{figure}

From the figure we see that the relaxation times seem to show an
apparent divergence upon approaching the wall. We have found that
at low temperatures this strong increase can be described well by
the functional form $\tau(q,\rho,T) \propto \exp[\Delta_q/(\rho_{\rm
p}-\rho)]$, where $\rho_{\rm p}$ is the penetration radius introduced
above, i.e. it is not a fit parameter. To demonstrate the quality of
this fit we plot in Fig.~\ref{fig4} the relaxation time in a logarithmic
way as a function of $(\rho_{\rm p}-\rho)^{-1}$ and we see that at low
temperatures the resulting fit is indeed very good (bold lines). With
increasing temperature we see the expected deviations at small $\rho$,
since there the relaxation times have to approach the bulk values,
i.e. they have to become independent of $\rho$. Nevertheless, even at
intermediate and moderately high temperatures the fit describes the data
well for large values of $\rho$. Note that the value of the parameter
$\Delta_q=7.8\pm0.2$ is independent of $T$, at least within the accuracy
of our data. Thus close to the wall the only effect of a decrease in
temperature is an increase of the prefactor of $\exp[\Delta_q/(\rho_{\rm
p}-\rho)]$. We have found that this temperature dependence is independent
of the wave-vector $q$ or the type of particle considered (A or B).
At low temperatures we can therefore write the whole temperature and
$\rho$-dependence of the relaxation time as

\begin{equation}
\tau(q,\rho,T) =C(q) f(T)  \exp[\Delta_q/(\rho_{\rm p}-\rho)] \, \, ,
\label{eq2}
\end{equation}

\noindent
where the function $f(T)$ depends strongly on temperature. We have found
that this dependence is compatible with an Arrhenius law, in agreement
with the experimental findings of Ref.~\cite{kremer_1999}, but also with a
power-law of the form $(T-T_c)^{-\gamma}$, a functional form suggested
by mode-coupling theory~\cite{goetze_1999}. The value of the critical
temperature $T_c$ is around $0.39\pm 0.02$, i.e. is significantly less
than the one found in the bulk which is 0.435~\cite{kob_1995}. Also the
value for the exponent, $\gamma=4.7 \pm 0.5$, is much larger than the bulk
value 2.35~\cite{kob_1995}. Whether or not the dynamics of the particles
in the confined geometry can be described by mode-coupling theory as
well as it is the case in the bulk is the focus of current investigations.

Before we conclude we briefly comment to what extent these results can be
generalized to other systems. It might be expected, e.g., that the strong
slowing down of the dynamics close to the wall is related to the fact
that the particles of the wall do not move at all, i.e. that they are at
zero temperature. However, we have repeated the present calculations also
for a system in which the particles in the wall behaved like interacting
Einstein oscillators and found that the dynamics is qualitatively the
same, i.e. also in this case a strong slowing down close to the wall
is found.  Furthermore we expect that the same type of slowing down is
observed also in a slit geometry, since the fact that the wall is curved
is probably irrelevant.

In summary we can say that by using a wall which does not affect the
static properties of the system we have been able to study the effect
of the confinement onto the dynamics of a simple liquid in a very clean
way. We find that the presence of the wall leads to a strong slowing
down of this dynamics and that the time correlation functions are much
stronger stretched in the confined system than in the bulk. By measuring
the dynamics of the particles as a function of their distance from the
wall, we find that the relaxation times show a very strong increase when
this distance is small. Therefore we are able to identify the reason
why the relaxation of the whole system is so stretched. Note that this
increase is a smooth function of the distance from the wall, i.e. we do
not see any evidence that the layer of particles closest to the wall
is decoupling dynamically from the rest of the liquid, as it has been
suggested before~\cite{schueller_1994,pissis_1998}. 
In the future it will be of interest to
see whether this type of simulation is also useful to identify a growing
length scale in supercooled liquids. Work in this direction is currently
done by Parisi and coworkers~\cite{parisi}.

Acknowledgement: We gratefully acknowledge the financial support by DFG
under SFB 262 and the NIC in J\"ulich.

\newpage

\begin{references}
\bibitem{footnote}
Note that we have determined $F_{\rm s}({\bf q},\rho,t)$ only for $\rho \geq
0.5$, since for smaller values of $\rho$ the statistics was not
satisfactory due to the small phase space.
\end{references}
\end{document}